\documentclass[11pt]{revtex4}
\usepackage{amssymb}
\usepackage{enumerate}
\usepackage{amsmath,bm}
\usepackage[latin1]{inputenc}

\begin{document}
\title{INFORMATION-THEORETIC PROPERTIES OF THE HALF-LINE COULOMB POTENTIAL}

\author{J.\ J.\ Omiste}

\affiliation{Instituto Carlos I de F\'{\i}sica Te\'orica y
Computacional and Departamento de F\'{\i}sica Atómica, Molecular y Nuclear, Universidad de Granada}

\author{R.\ J.\ Y\'añez}

\affiliation{Instituto Carlos I de F\'{\i}sica Te\'orica y
Computacional and Departamento de Matemática Aplicada, Universidad de Granada}

\author{J.\ S.\ Dehesa}

\affiliation{Instituto Carlos I de F\'{i}sica Te\'orica y
Computacional and Departamento de F\'{\i}sica Atómica, Molecular y Nuclear, Universidad de Granada}

\date{\today}

\begin{abstract}
The half-line one-dimensional Coulomb potential is possibly the
simplest D-dimen\-sional model with physical solutions which has been
proved to be successful to describe the behaviour of Rydberg atoms in
external fields and the dynamics of surface-state electrons in liquid
helium, with potential applications in constructing analog quantum
computers and other fields. Here, we investigate the spreading and
uncertaintylike properties for the ground and excited states of this
system by means of the logarithmic measure and the
information-theoretic lengths of Renyi, Shannon and Fisher types; so,
far beyond the Heisenberg measure. In particular, the Fisher length
(which is a local quantity of internal disorder) is shown to be the
proper measure of uncertainty for our system in both position and
momentum spaces. Moreover the position Fisher length of a given
physical state turns out to be not only directly proportional to the
number of nodes of its associated wavefunction, but also it follows a
square-root  energy law.
\end{abstract}

\maketitle



\section{Introduction}

The one-dimensional (1D) hydrogenic atom is the simplest known model proved to be very useful to describe Rydberg atoms \cite{cas,may} to investigate the microwave ionization of highly excited atoms \cite{leo,sto}, to study the behavior of atoms irradiated by strong laser fields \cite{pen} and the dynamics of surface-state electrons in liquid helium \cite{jen,nie}. The latter has allowed this system to be considered as potentially useful in constructing analog quantum computers \cite{pla,dyk} and, particularly, as a possible candidate for a 2-qubit quantum gate \cite{jak}. Moreover, this system has been used (i) to analyze the main features of revival and fractional revival phenomena \cite{vei} and the capabilities of functional-analytic approaches for the construction of quantum hamiltonians with non-standard dimensionalities \cite{fis}, and (ii) to simulate the interior of giant planets in planetary physics \cite{gui}.

The position-space wavefunctions for the 1D hydrogenic atom, $\Psi(x,t)=\psi(x)\exp(-i\frac{Et}{\hbar})$, are determined by the eigensolutions $(\psi,E)$ of its Hamiltonian:
\begin{equation}
H=-\cfrac{\hbar^2}{2m}\cfrac{d^2}{dx^2}+
	\begin{cases}
	-\cfrac{Ze^2}{x}, &  \hspace{0.5cm} x>0 \\
	\infty, &  \hspace{0.5cm} x<0
	\end{cases}
\end{equation}
which are quite well-known \cite{nie}. In atomic units ($\hbar=m=e=1$) they are given by
\begin{equation}
E_n=-\cfrac{Z^2}{2n^2}; \hspace{1cm} n=1, 2, ...
\label{eqn:energy}
\end{equation}
for the energetic eigenvalues, and
\begin{equation}
\psi_n(x)=\left(\cfrac{Z}{n^3}\right)^{1/2}z_n e^{-\frac{z_n}{2}}L_{n-1}^{(1)}(z_n)= \left(\cfrac{Z}{n^3}\right)z_n^{1/2}\sqrt{\omega_1(z_n)}L_{n-1}^{(1)}(z_n)
\label{eqn:psi}
\end{equation}
for the corresponding eigenfunctions, where $z_n=\frac{2Z}{n}x$ and $L_k^{(\alpha)}(t)$ denotes the Laguerre polynomials which are orthogonal with respect to the weight function $\omega_\alpha (x)\equiv x^\alpha e^{-x}$ on the interval $[0,\infty)$; that is
\begin{equation}
\int_0^\infty L_n^{(\alpha)}L_m^{(\alpha)}(x)\omega_\alpha(x)dx=\frac{\Gamma(n+\alpha+1)}{n!}
\end{equation}

Not so familiar are the wavefunctions in momentum space. Nevertheless, the Fourier transform of $\psi_n(x)$ can be done straightforwardly to obtain the expression and it is a particular case of what is done in \cite{ave}:
\begin{equation}
\phi_n(p)=\left(\cfrac{2n}{\pi Z}\right)^{1/2}\cfrac{1}{\left(\frac{np}{Z}\right)^2+1}\exp\left(-2in\arctan\left(\frac{np}{Z}\right)\right)
\label{eqn:phi}
\end{equation}  

Then, the quantum-mechanical probability densities of the system in the two reciprocal spaces are the squares of the corresponding position and momentum eigenfunctions given by Eqs. (\ref{eqn:psi}) and (\ref{eqn:phi}), respectively. We have
\begin{equation}
\rho_n(x)=\cfrac{Z}{n^3}z_n^2e^{-z_n}[L_{n-1}^{(1)}(z_n)]^2;\hspace{1cm} x>0
\label{eqn:rho}
\end{equation}  
in position space, and
\begin{equation}
\gamma_n(p)=\cfrac{2n}{\pi Z}\frac{1}{\left(\frac{n^2p^2}{Z^2}+1\right)^2};\hspace{1cm} -\infty<p<+\infty
\label{eqn:gamma}
\end{equation}  
in momentum space.

Here we investigate the spreading and uncertainty properties of the 1D hydrogen atom in the light of the information theory of quantum systems. For this purpose we use not only the standard deviation or Heisenberg uncertainty measure and the ordinary moments of the position and momentum densities of the system, but also their entropic moments and their associated information-theoretic lengths.
The structure of this work is the following. First, in Section 2 we define the spreading and uncertainty measures used in this work and we give their main properties and interpretations. Then, in Section 3, we calculate the position and momentum power and logarithmic moments as well as their associated uncertainty measures for the 1D hydrogenic system. In Section 4 we evaluate the position and momentum information-theoretic lengths and their associated uncertainty measures. Finally, some conclusions are given.

\section{Spreading and uncertainty measures: concept and properties}

The spreading of a probability density $f(x)$ on its domain of definition $\Delta=(a,b)$ is known to be measured by its moments around a particular point $x_0\in\Delta$ defined by
\begin{equation}
\mu_k(x_0)=\int_a^b(x-x_0)^k f(x)dx\equiv \langle(x-x_0)^k\rangle,
\end{equation}
for $k=0,1,2...$. The best known moments are the moments around the origin, $\mu_k'=\langle x^k\rangle$. There exists, however, another kind of quantities which measure the spreading of the probability cloud in a qualitatively different manner: the frequency moments $W_k$, also called entropic moments because they are closely connected to various entropy measures in information theory. They are defined for the continuous univariate density $f(x)$ \cite{yul,si1,ken} as
\begin{equation} 
W_q[\rho_n]=\int_a^b[\rho_n(x)]^q dx\equiv \langle[\rho_n(x)]^{q-1}\rangle;\quad q>0
\label{eqn:nuk}
\end{equation}

These moments have two main characteristics which make them to be, at times, much better probability estimators than the previous moments \cite{si1,si2}. First, the frequency moments measure the extent in which the probability is spread without respect to any specific point of its support interval $\Delta$. Moreover, they are fairly efficient in the range where the ordinary moments are fairly inefficient \cite{she}. A recent summary about these quantities can be seen in \cite{rom}.

In quantum theory the position and momentum uncertainty measures are often related to the ordinary and/or frequency moments of the corresponding quantum-mechanical probability density in position and momentum spaces, respectively. The most popular one, mainly because of mathematical convenience, is the Heisenberg measure which is defined by the standard or root-mean-square deviation of the quantum probability density $\rho_n(x)$ characterized by the quantum number $n$:
\begin{equation}
(\Delta x)_n=\sqrt{\langle x^2\rangle_n-\langle x\rangle_n^2},
\label{eqn:errx}
\end{equation}
where $\langle x^2\rangle_n$ and $\langle x\rangle_n^2$ denote the first and second order moments around the origin of $\rho_n(x)$. This quantity has various interesting properties. First, it is a direct measure of spreading in the sense of having the same units as the variable. Second, it is invariant under translation, i.e. independent of the location of the density function. Third, it scales linearly, i.e. $\Delta y=\lambda \Delta x$ for $y=\lambda x$. And, finally, it vanishes as the density approaches a Dirac delta function, i.e. in the limit that $x$ tends towards a given definite value. However the Heisenberg measure, although good enough for Gaussian-like densities, is not generally an useful measure of uncertainty; at times, it is misleading or not defined \cite{uff, ha1}. It does not measure the extent to which the distribution is in fact concentrated but rather the separation of the region(s) of concentration from a particular point of the distribution (namely the centroid or mean value). Moreover, it is undefined, e. g., for the Cauchy-Lorentz ($\rho(x)=\pi^{-1}/(1+x^2)$) and the sink-squared ($\rho(x)=\pi^{-1}(\frac{\sin(x)}{x})^2$) distributions.

Recently, various information-theoretic-based uncertainty measures have been proposed \cite{ha1,ha2} which take care of these defects but keep the nice properties of the Heisenberg measure mentioned above. They are the Renyi, Heller or Onicescu, Shannon and Fisher information-theoretic lengths \cite{ha1,ha2}. They are defined by
\begin{equation}
L_q^R[\rho_n] = \exp(R_q[\rho_n])=\langle[\rho_n(x)]^{q-1}\rangle^{-\frac{1}{q-1}};\quad q>1,\hspace{0.5cm}
\label{eqn:renyi}
\end{equation}
for the Renyi length of order $q$ (which is the exponential of the Renyi entropy $R_q[\rho_n]$),
\begin{equation}
L^O[\rho_n]=\langle \rho_n(x)\rangle^{-1}
\label{eqn:heller}
\end{equation}
for the Onicescu length \cite{hel} (also called collision or Heller length, inverse disequilibrium and participation ratio in other contexts),
\begin{equation}
H[\rho_n]=\exp({S[\rho_n]});\hspace{1cm} S[\rho_n] \equiv -\int_a^b dx \rho_n(x)\ln\rho_n(x)
\label{eqn:shannon}
\end{equation}
for the Shannon length, and
\begin{equation}
(\delta x)_n=\cfrac{1}{\sqrt{F[\rho_n]}};\hspace{1cm} F[\rho_n]=\int_a^b dx \cfrac{[\rho_n'(x)]^2}{\rho_n(x)}
\label{eqn:fisherpos}
\end{equation}
for the Fisher length, where $F[\rho_n]$ denotes the Fisher information of $\rho_n(x)$. Remark that the Shannon and Onicescu lengths correspond to the cases $q=0$ and $1$, respectively, of the Renyi lengths.

The Renyi, Onicescu and Shannon lengths are, as the Heisenberg measure, uncertainty measures of global character because they are quadratic or logarithmic functionals of the probability density. On the contrary, the Fisher length is a local measure of uncertainty because it is a functional of the derivative of the density. So, it measures the pointwise concentration of the probability cloud along its support domain and quantifies its gradient content, providing a quantitative estimation of the oscillatory character of the density. Moreover, the Fisher length measures the bias to particular points of the interval, i.e. it gives the degree of local disorder. Let us also mention that each of these measures satisfies an uncertainty relation: see \cite{ha1} for the Heisenberg, Shannon and Renyi cases, and \cite{de1,de2} for the Fisher case. Moreover, they fulfil the inequalities \cite{ha1}
\begin{equation}
(\delta x)_n\leq (\Delta x)_n \hspace{1cm} \mbox{ and } \hspace{1cm} H[\rho_n]\leq (2\pi e)^{1/2}(\Delta x)_n,
\label{eqn:ineq}
\end{equation}
where the equality is reached if and only if the density $\rho_n(x)$ is Gaussian. These inequalities illustrate that the Shannon and Fisher lengths, as measures of uncertainty, have a range of applicability much wider than the standard deviation.

\section{Power and logarithmic moments of the model. The Heisenberg and logarithmic uncertainty measures}
\subsection{Position space}
Let us calculate the moments around the origin $\langle x^k\rangle$ and the logarithmic moments $\langle(\log x)^k\rangle$ of the 1D hydrogenic atom defined by 
\begin{equation}
\langle x^k \rangle_n=\int_0^\infty x^k\rho_n(x)dx,
\label{eqn:xk}
\end{equation}
\begin{equation}
\langle(\log x)^k\rangle_n=\int_0^\infty (\log x)^k\rho_n(x)dx,
\label{eqn:logxk}
\end{equation}
respectively, where $\rho_n(x)$ denotes the position-space probability density (\ref{eqn:rho}) of the system. Then, we calculate the corresponding quantities in momentum space; and, finally, we determine the associated Heisenberg and logarithmic uncertainty products.

>From Eqs. (\ref{eqn:rho}) and (\ref{eqn:xk}) one has that the power moments of order $k$ are given by
\begin{align}
\langle x^k \rangle_n &= \cfrac{Z}{n^3}\int_0^\infty x^k z_n^2 e^{-z_n}[L_{n-1}^{(1)}(z_n)]^2dx \\
\nonumber
&= \cfrac{n^{k-2}}{Z^k2^{k+1}}\int_0^\infty t^{k+1}\omega_1(t)[L_{n-1}^{(1)}(t)]^2dt
\label{eqn:xkint}
\end{align}
Then, taking into account the relation
\begin{equation}
\int_0^\infty \omega_{\alpha}(x) x^{s-\alpha} [L_n^{(\alpha)}(x)]^2dx=s!\sum_{r=0}^n (-1)^r \binom{s-\alpha}{n-r}^2\binom{-s-1}{r},
\label{eqn:xkint2}
\end{equation}
we find that (i) the only power moments with negative orders are $\langle x^{-1}\rangle$ and $\langle x^{-2}\rangle$, with the following values
\begin{align}
\langle x^{-1}\rangle_n &= \cfrac{Z}{n^2}\\
\langle x^{-2}\rangle_n &= \cfrac{2Z^2}{n^3}
\end{align}
and (ii) the power moments with positive order $k$ are given by
\begin{align}
\langle x^{k}\rangle_n &=
\cfrac{n^{k-1}}{2^{k+1}Z^{k}}\sum_{i=0}^{k+1}\cfrac{(-1)^i(n+k-i)!}{(n-1-i)!}\binom{n}{i}\binom{n+k-i}{k+1-i},\hspace{.5cm}
k \in \mathbb{N}_0  \nonumber \\
&= \cfrac{n^{k-2}}{2^{k+1}Z^{k}}\sum_{i=0}^{n-1}	\binom{k+1}{n-i-1}^2\cfrac{(k+i+2)!}{i!},\hspace{.5cm} k \in \mathbb{R}
	\label{eqn:xksumn}
\end{align}

For the three lowest orders we find that $\langle x^0\rangle=1$, as it must be, and
\begin{align}
\langle x \rangle_n &= \cfrac{3}{2Z}n^2\\
\langle x^2 \rangle_n &= \cfrac{n^2}{2Z^2}(5n^2+1)
\end{align}
respectively. Then, the position-space Heisenberg uncertainty measure
is given by the root-mean-square or standard deviation
\begin{equation}
(\Delta x)_n=\sqrt{\langle x^2 \rangle_n-\langle x \rangle_n^2}=\frac{n}{Z}\sqrt{n^2+2}
\label{eqn:deltax}
\end{equation}
Working similarly it is possible to calculate the logarithmic uncertainty measure defined by
\begin{equation}
(\Delta \log x)_n=\sqrt{\langle (\log x)^2 \rangle_n-\langle \log x \rangle_n^2},
\end{equation}
where the involved logarithmic moments can be found from Eq. (\ref{eqn:xksumn}) and the limiting relation
\begin{equation}
\langle(\log x)^m\rangle=\lim_{\alpha\rightarrow 0}\frac{\partial^m\langle x^\alpha\rangle}{\partial \alpha^m}; \hspace{1cm} m=1,2...
\end{equation}
Then we have
\begin{align}
\langle\log x\rangle_n &= \log\frac{en}{2Z}-\frac{1}{2n}+\psi(n+1),\\
\langle(\log x)^2\rangle_n &= \left(\log
\frac{en}{2Z}\right)^2+\psi(n+1)\left[\psi(n+1)+2\log
\frac{en}{2Z}-\frac{1}{n}\right]-\frac{1}{n}\log \frac{en}{2Z} \nonumber \\
&+\psi'(n+1)+\frac{(1-\delta_{n,1})(1-\delta_{n,2})}{n^2}\sum_{i=0}^{n-3}\cfrac{(i+2)(i+1)}{(n-i-2)^2(n-i-1)^2},
\end{align}
where $\psi(x)=\frac{\Gamma'(x)}{\Gamma(x)}$ is the digamma function and $\delta_{n,k}$ is the Kronecker delta. So that the logarithmic uncertainty measure becomes
\begin{equation}
(\Delta \log x)_n= \begin{cases}
           \left(\psi'(n+1)+\cfrac{1}{4n^2}\right)^{1/2}, &
           \hspace{-5cm} n= 1, 2
\\ \left(\psi'(n+1)+\cfrac{1}{n^2}{\displaystyle \sum_{i=0}^{n-3}}\cfrac{(i+2)(i+1)}{(n-i-2)^2(n-i-1)^2}+\cfrac{1}{4n^2}\right)^{1/2}, & n>2    \\
  
\end{cases} 
\label{eqn:deltalogx}
\end{equation}
which, contrary to the Heisenberg uncertainty measure $(\Delta x)_n$ given by Eq. (\ref{eqn:deltax}), does not depend on Z.

\subsection{Momentum space}

Let us now calculate the momentum power moments which are defined by
\begin{equation}
\langle p^k \rangle=\int_{-\infty}^{+\infty}p^k\gamma_n(p)dp=\frac{2n}{\pi Z}\int_{-\infty}^{+\infty}\cfrac{p^k}{(\frac{n^2p^2}{Z^2}+1)^2}dp,
\end{equation}
where $\gamma_n(p)$ is the probability density of the system in momentum space given by Eq. (\ref{eqn:gamma}). From this expression it is clear that
\begin{equation}
\langle p^k\rangle=0, \hspace{1cm} k=1,3,5...
\end{equation}
and for $k$ even with the change of variable $p\rightarrow t:t=\frac{np}{Z}$ and the relation
\begin{equation}
\nonumber
\int_0^\infty\cfrac{x^adx}{(x^2+1)^b}=\frac{1}{2}\cfrac{\Gamma\left(\frac{1+a}{2}\right)\Gamma\left(b-\frac{a+1}{2}\right)}{\Gamma(b)}, \hspace{1cm} 2b>a+1
\label{eqn:intbessel}
\end{equation}
we finally find that
\begin{equation}
 \langle p^0\rangle=1, \hspace{.5 cm}\langle p^2\rangle=\cfrac{Z^2}{n^2}
 \label{eqn:p2exp}
\end{equation}
and the Heisenberg uncertainty measure in momentum space has the value
\begin{equation}
(\Delta p)_n=\sqrt{\langle p^2 \rangle_n}=\frac{Z}{n}
\label{eqn:errp}
\end{equation}

So, the position-momentum Heisenberg uncertainty product has, according to Eqs. (\ref{eqn:errx}) and (\ref{eqn:errp}), the value
\begin{equation}
(\Delta x)_n(\Delta p)_n=\sqrt{n^2+2};\hspace{1cm} n=1, 2,...
\label{eqn:heisenberg}
\end{equation}
which is not only greater than 1/2 as the uncertainty principle states, but also it does not depend on the strength Z of the potential as recently predicted by S. H. Patil et al \cite{pat} for a general class of potentials which includes the present one.

For the sake of completeness let us also calculate the absolute power and logarithmic moments. In particular we have for the power ones the values
\begin{equation}
\langle |p|\rangle = \cfrac{2Z}{n\pi}, \hspace{.5cm} \langle |p|^2\rangle = \cfrac{Z^2}{n^2}
\end{equation}
so that $\Delta |p|=\frac{Z}{n}\sqrt{1-\frac{4}{\pi^2}}$, and for the absolute logarithmic moments the values
\begin{equation}
\langle\log |p|\rangle_n = -\left(1+\log\frac{n}{Z}\right),\hspace{.5cm} \langle(\log |p|)^2\rangle_n = \frac{\pi^2}{4}+\log\frac{n}{Z}\left(2+\log\frac{n}{Z}\right)
\end{equation}
Then the logarithmic uncertainty measure yields:
\begin{equation}
(\Delta\log |p|)_n=\sqrt{\frac{\pi^2}{4}-1}
\label{eqn:deltalogp}
\end{equation}
which does not depend on the value of $Z$. Finally, it is worthy to remark that the logaritmic uncertainty product has the following value
\begin{multline}
(\Delta \log x)_n(\Delta \log |p|)_n=\\ \left[\frac{\pi^2-4}{4}\left(\psi'(n+1)+\frac{1}{n^2}\sum_{i=0}^{n-3}\cfrac{(i+2)(i+1)}{(n-i-2)^2(n-i-1)^2}+\cfrac{1}{4n^2}\right)\right]^{1/2},
\end{multline}
where Eqs. (\ref{eqn:deltalogx}) and (\ref{eqn:deltalogp}) were taken into account.
\section{Information-theoretic lengths and their associated uncertainty measures}

\subsection{Position space}
Let us here study the position Fisher, Renyi and Shannon information-theoretic lengths of the 1D hydrogenic model, which are defined by Eqs. (\ref{eqn:fisherpos}), (\ref{eqn:renyi}) and (\ref{eqn:shannon}) respectively.

\subsubsection{\emph{Fisher length $(\delta x)_n$}}
	
	From Eqs. (\ref{eqn:fisherpos}) and (\ref{eqn:rho}) one has that the Fisher information $F[\rho_n]$ of the model is:
	\begin{align}
	\nonumber
	F[\rho_n]&=\int_0^\infty \cfrac{[\rho_n'(x)]^2}{\rho_n(x)}dx=-4\int_0^\infty \psi_n(x) \psi_n''(x)dx\\	
	&= 4\langle p^2\rangle_n=\cfrac{4Z^2}{n^2},
	\label{eqn:fn}
	\end{align}	
	where we have made an integration by parts for the second equality, and then used the Schrödinger equation which allows us to write $\psi''(x)=-p^2\psi(x)$. Finally the value (\ref{eqn:p2exp}) for the expectation value of $\langle p^2 \rangle_n$ was used.
	Then, from (\ref{eqn:fisherpos}) and (\ref{eqn:fn}) one has that the Fisher length is 
	\begin{equation}
	(\delta x)_n=\cfrac{1}{\sqrt{F_n}}=\cfrac{n}{2Z}
	\label{eqn:fisherlengthx}
	\end{equation}
	for a quantum state characterized by the principal quantum
	number $n$. Then, the Fisher length of a physical state is
	directly proportional to the number of nodes of its
	corresponding wavefunction. Moreover, from  Eqs. (\ref{eqn:deltalogx}) and (\ref{eqn:fisherlengthx}) 
	\begin{equation}
	(\delta x)_n=\frac{(\Delta x)_n}{2\sqrt{n^2+2}} \le \frac{(\Delta x)_n}{2\sqrt{3}}\simeq 0.288 (\Delta x)_n
	\label{eqn:shannonlx}
	\end{equation}
	which shows that the Fisher length is always much less than
	the standard deviation in position space, what it is in
	accordance to the general inequality (\ref{eqn:ineq}),
	indicating that the Fisher length is a measure of uncertainty
	much more appropriate than the Heisenberg measure for our
	system. Finally, from Eqs. (\ref{eqn:energy}) and (\ref{eqn:fisherlengthx}) one finds
	that $(\delta x)_n=\left(8 | E_n|\right)^{1/2}$ which shows
	that the position Fisher length of a physical state has a
	square-root dependence on its corresponding energy.
	
\subsubsection{\emph{Renyi lenghts $L_q^R[\rho_n]$}}
	
	According to Eq. (\ref{eqn:nuk}), these quantities are controlled by the entropic moments
	\begin{align}
	\nonumber
	W_{q}[\rho_n]&=\langle
[\rho_n(x)]^{q-1}\rangle=\int_0^\infty [\rho_n(x)]^q dx\\ \nonumber
	&= \left(\frac{Z}{n^3}\right)^q\cfrac{n}{2Z}\int_0^\infty [t e^{-\frac{t}{2}}L_{n-1}^{(1)}(t)]^{2q}dt\\
	&= \left(\frac{Z}{n^3}\right)^q\cfrac{n}{2Z}\int_0^\infty t^q [\omega_1(t)L_{n-1}^{(1)}(t)^2]^{q}dt,\hspace{1cm} q>1
	\label{eqn:nupos}
	\end{align}
	where we have used Eq. (\ref{eqn:rho}) in the third equality. Remark that the kernel of the integral involves the Rakhmanov density of the Laguerre polynomials $L_{n-1}^{(1)}(t)$; i. e. $\omega_1(t)[L_{n-1}^{(1)}(t)]^2$.
	
	Except for the trivial case $q=1$ (then, $W_0=1$), the closed expressions for the Krein-like functional of the Laguerre polynomial $L_{n-1}^{(1)}(t)$ involved in the calculation of the qth-entropic moment of an arbitrary physical state (i. e., for any value of n) has not yet been found; not even for the case $q=2$, where
	\begin{align}
	\nonumber
	W_2[\rho_n]&=\langle \rho_n(x) \rangle=\frac{Z}{2n^5}\int_0^\infty [t e^{-\frac{t}{2}}L_{n-1}^{(1)}(t)]^4dt\\
	&= \frac{Z}{2n^5}\int_0^\infty t^2\{\omega_1(t)[L_{n-1}^{(1)}(t)]^2\}^2 dt
	\label{eqn:nuq2pos}
	\end{align}
	
	However, for the ground state (whose density $\rho_{g. s.}(\bar{r})=\rho_{1}(\bar{r})$) the involved Laguerre polynomial reduces to a constant, and its associated entropic moments have the values	
	\begin{equation}	
	W_{q}[\rho_{g. s.}]= \frac{Z^{q-1}}{2}\int_0^\infty t^q [\omega_1(t)]^{2q}dt= \frac{Z^{q-1}}{2}\frac{\Gamma(2q+1)}{q^{2q+1}}
	\end{equation}
	
	Then, the Renyi lengths for the ground state of the system are	
	\begin{equation}
	L_{q+1}^R[\rho_{g. s.}]=(W_q[\rho_{g. s.}])^{-\frac{1}{q}}= \frac{(q+1)^{2+3/q}}{Z}\left(\cfrac{2}{\Gamma(2q+3)}\right)^{1/q};\qquad q>0
	\end{equation}
	and the Onicescu length ($q=1$) has the value
	\begin{equation}
	L^O[\rho_{g. s.}]=L_2^R[\rho_{g. s.}]=\frac{8}{3Z}
	\end{equation}
	
	
\subsubsection{\emph{The Shannon length $H[\rho_n]$}}
	
	To calculate this quantity we need first to compute the Shannon entropy $S_\rho$ of the model, which is given by Eq. (\ref{eqn:shannon}). That is,
	\begin{equation}
	S[\rho_n]=-\int_0^\infty dx \rho_n(x) \log \rho_n(x)=-\int_0^\infty dx |\psi_n(x)|^2 \log |\psi_n(x)|^2
	\label{eqn:srho}
	\end{equation}
	Then, taking into account Eq. (\ref{eqn:psi}) and with the change of variable $x\rightarrow t=\frac{2Z}{n}x$, one can express
	\begin{equation}
	S[\rho_n]=-\frac{1}{2n^2}\left[\log \left(\frac{Z}{n^3}\right)I_1[L_{n-1}^{(1)}]+I_2[L_{n-1}^{(1)}]+E_1[L_{n-1}^{(1)}]\right]
	\label{eqn:shannonrhox}
	\end{equation}	
	where the symbol $I_i[L_{n-1}^{(1)}]$ denotes the integrals
	\begin{align}
	\nonumber
	I_1[L_{n-1}^{(1)}]&=\int_0^\infty t^2 e^{-t}[L_{n-1}^{(1)}(t)]^2\\
	\nonumber
	&= \int_0^\infty t \omega_1(t)[L_{n-1}^{(1)}]^2dt\\
	&= 2n^2,
	\label{eqn:i1}
	\end{align}
	\begin{align}
	\nonumber
	I_2[L_{n-1}^{(1)}]&=\int_0^\infty t^2 e^{-t}\log(t^2 e^{-t})[L_{n-1}^{(1)}(t)]^2\\	
	&= 4n^2+4n^2\psi(n+1)-6n^3-2n
	\label{eqn:i2}
	\end{align}
	using the Eqs. (6) and (30) from \cite{san} to obtain (\ref{eqn:i2}). The integral $E_1[L_{n-1}^{(1)}]$ is given by
	\begin{equation}
	E_1[L_{n-1}^{(1)}]= \int_0^\infty t \omega_1(t)[L_{n-1}^{(1)}(t)]^2 \log [L_{n-1}^{(1)}(t)]^2dt	
	\label{eqn:eshannonrhox}
	\end{equation}	
	which has not yet been exactly solved for any generic $n$, except for its lowest and highest values. For the ground state this entropic integral vanishes so that the Shannon entropy has the value $S[\rho_{g. s.}]=2\gamma-\log Z$, where $\gamma\simeq 0.577$ is the Euler-Mascheroni constant, and the associated Shannon length is
	\begin{equation}
	H[\rho_{g. s.}]=\frac{e^{2\gamma}}{Z}
	\label{eqn:slxground}
	\end{equation}
	In the asymptotic region (i. e., for large $n$) one has from \cite{dya} that
	\begin{equation}
	E_1[L_{n-1}^{(1)}]= 2n^2(3n-\log n-\log 2\pi+o(1)),
	\label{eqn:e1}
	\end{equation}
	Then, we finally obtain from (\ref{eqn:srho})-(\ref{eqn:e1}) that the Shannon entropy of the 1D-hydrogenic Rydberg states, whose quantum probability density $\rho_{Ry}(\bar{r})=\rho_{n}(\bar{r})$ with large $n$, has the value
	\begin{equation}
	S[\rho_{Ry}]=\log \left(\frac{2\pi n^2}{Ze^2}\right)+o(1)
	\end{equation}
	and the corresponding Shannon length is given by
	\begin{equation}
	H[\rho_{Ry}]=e^{S[\rho_{Ry}]}\simeq\frac{2\pi n^2}{Ze^2}
	\label{eqn:shannonlenghtasym}
	\end{equation}
	Finally, taken into account Eqs. (\ref{eqn:deltax}) and (\ref{eqn:shannonlenghtasym}), we observe that
	\begin{equation}
	H[\rho_{Ry}]\simeq \frac{2\pi}{e^2}(\Delta x)_{Ry}\simeq 0.85(\Delta x)_{Ry}
	\end{equation}
	for very excited states, what it is in agreement with the general inequality (\ref{eqn:ineq}). 
	
	In summary, although we have not been able to analytically evaluate the Shannon length for arbitrary  states because of the difficulties to calculate the involved entropic funtional $E_1[L_{n-1}^{(1)}]$, we have found its values for the energetically extremal states of the 1D hydrogenic spectrum: the ground and Rydberg states. The comparison of these values with those of the Heisenberg, Onicescu and Fisher lengths yields the inequalities
	\begin{equation}
	(\delta x)_{g. s.}<(\Delta x)_{g. s.}<L^O[\rho_{g. s.}]<H[\rho_{g. s.}]
	\end{equation}
	for the ground state, and
	\begin{equation}
	(\delta x)_{Ry}<H[\rho_{Ry}]<(\Delta x)_{Ry}
	\end{equation}
	for the Rydberg states. They show, together with relation (\ref{eqn:shannonlx}), that the Fisher length is the proper position measure of uncertainty of our system.

\subsection {Momentum space}
Now we will compute the information-theoretic lengths in momentum space for this model.

\subsubsection{\emph{Fisher length $(\delta p)_n$}}
	
	Similarly to the position case (\ref{eqn:fisherpos}), the momentum Fisher length is defined by
	\begin{equation}
	(\delta p)_n=\frac{1}{\sqrt{F[\gamma_n]}}
	\label{eqn:fisherlengthmom}
	\end{equation}
	where the Fisher information $F[\gamma_n]$ of the momentum density $\gamma_n(p)$ given by Eq. (\ref{eqn:gamma}) is
	\begin{align}
	\nonumber
	F[\gamma_n] &= \int_{-\infty}^{\infty}dp\cfrac{[\gamma_n'(p)]^2}{\gamma_n(p)}\\
	\nonumber
	&= \frac{2n}{Z\pi}\int_{-\infty}^{\infty}\left(\cfrac{-4\frac{n^2p}{Z^2}}{\left(\frac{n^2p^2}{Z^2}+1\right)^3}\right)^2\left(\frac{n^2p^2}{Z^2}+1\right)^2dp\\
	&=\frac{32n^2}{\pi Z^2}\int_{-\infty}^\infty \frac{t^2}{(t^2+1)^4}dt=\frac{2n^2}{Z^2}
	\label{eqn:fishermom}
	\end{align}
	where we have made the change $t\rightarrow \frac{np}{Z}$ and used Eq. (\ref{eqn:intbessel}). So the Fisher length in momentum space is
	\begin{equation}
	(\delta p)_n=\frac{Z}{\sqrt{2}n}
	\label{eqn:fisherlengthp}
	\end{equation}
	The comparison of Eqs. (\ref{eqn:errp}) and (\ref{eqn:fisherlengthp}) shows that $(\delta p)_n=\frac{1}{\sqrt{2}}(\Delta p)_n\simeq 0.707 (\Delta p)_n$, so that the Fisher length is less than the standard deviation in momentum space. Moreover, the position and momentum Fisher lengths fulfil the following uncertainty equality
	\begin{equation}
	(\delta x)_n(\delta p)_n=\frac{1}{2\sqrt{2}}\simeq 0.354,
	\label{eqn:fisherlengthsineq}
	\end{equation}
	where Eqs. (\ref{eqn:fisherlengthx}) and
(\ref{eqn:fisherlengthp}) have been taken into account. It is worthy
to remark that this Fisher uncertainty product does not depend neither
of $Z$ nor of the quantum number $n$.

\subsubsection{\emph{Renyi lengths $L_q^R[\gamma_n]$}}
	
	As we have already seen, the Renyi lengths are related to the entropic moments, as given by Eq. (\ref{eqn:renyi}). Moreover, we can calculate exactly in the momentum space
	\begin{align}
	\nonumber
	W_{q}[\gamma_n]&=\langle [\gamma_n(p)]^{q-1}\rangle=\int_{-\infty}^\infty [\gamma_n(p)]^qdp\\
	\nonumber
	&= \left(\frac{2n}{Z\pi}\right)^q\int_{-\infty}^\infty \cfrac{1}{\left(\frac{n^2p^2}{Z^2}+1\right)^{2q}}dp\\
	\nonumber
	&= \left(\frac{2}{\pi}\right)^q \left(\frac{n}{Z}\right)^{q-1}\int_{-\infty}^\infty \cfrac{dt}{\left(t^2+1\right)^{2q}}\\
	&= \left(\frac{n}{8\pi Z}\right)^{q-1}\binom{4q-3}{2q-1}
	\end{align}
	where we have used Eqs. (\ref{eqn:gamma}) and (\ref{eqn:intbessel}). Then, the Renyi lengths turn out to be
	\begin{equation}
L_{q+1}^R[\gamma_n]=\frac{8\pi Z}{n}\binom{4q+1}{2q+1}^{\frac{-1}{q}};\qquad q>0
	\label{eqn:renyimom}
	\end{equation}
	
	As we already know, the Onicescu length is a particular case of the Renyi length, according to Eq. (\ref{eqn:heller}). Using this in Eq. (\ref{eqn:renyimom}) we can obtain:
	\begin{equation}
	L^O[\gamma_n]=L_2^R[\gamma_n]=\frac{4\pi Z}{5 n}
	\label{eqn:hellermom}
	\end{equation}

\subsubsection{\emph{The Shannon length $H_n$}}
	
	First of all, we must compute the Shannon entropy $S[\gamma_n]$ given by (\ref{eqn:shannon}). This yields
	\begin{align}
	\nonumber
	S[\gamma_n]&=-\left(\frac{2n}{\pi Z}\right)\log\left(\frac{2n}{\pi Z}\right)\int_{-\infty}^\infty \cfrac{dp}{\left(\frac{n^2p^2}{Z^2}+1\right)^2}dp-\frac{2n}{\pi Z}\int_{-\infty}^\infty \cfrac{\log\left(\frac{n^2p^2}{Z^2}+1\right)^{-2}}{\left(\frac{n^2p^2}{Z^2}+1\right)^2}dp\\
	&\nonumber = \frac{8}{\pi}\int_0^\infty\cfrac{\log(t^2+1)}{(t^2+1)^2}dt+\log\frac{\pi Z}{2n}\\
	&= 2(\log 4 -1)+\log\frac{\pi Z}{2n}
	\end{align}	
	where we have used the following general expression	
	\begin{equation}
	\int_0^\infty \frac{\log(1+t^2)}{(1+t^2)^\alpha}dt=\frac{\sqrt{\pi}}{2}\frac{\Gamma\left(\alpha-\frac{1}{2}\right)}{\Gamma(\alpha)}[\psi\left(\alpha\right)-\psi\left(\alpha-1/2\right)],\hspace{1cm} \alpha>\frac{1}{2}
	\end{equation}
	And finally, using this result and Eq. (\ref{eqn:shannon}) we obtain
	\begin{equation}
	H[\gamma_n]=e^{S[\gamma_n]}=\frac{8\pi Z}{ne^2}
	\label{eqn:shannonlengthp}
	\end{equation}
	
	The comparison of Eqs. (\ref{eqn:energy}) and (\ref{eqn:fisherlengthp})-(\ref{eqn:shannonlengthp}) allows us to make various observations. First, the relation of the spreading measures of the momentum wavefunction with its corresponding energy is given by 	
	\begin{equation}
	|E_n|=\frac{1}{2}(\Delta p)_n^2=\frac{1}{2}(\delta p)_n^2=\frac{(8\pi)^2}{2}(L[\gamma_n])^2=\frac{1}{2}\left(\frac{e^2}{8\pi}\right)^2(H_n[\gamma_n])
	\end{equation}
	
	Second, the following chain of inequalities is fulfilled	
	\begin{equation}
	(\delta p)_n<(\Delta p)_n<H[\gamma_n]
	\end{equation}
	which illustrates, in particular, that the Fisher length is a measure of uncertainty more appropriate than the Heisenberg and Shannon length in the momentum space too.
	Moreover, the relations (\ref{eqn:slxground}), (\ref{eqn:shannonlenghtasym}) and (\ref{eqn:shannonlengthp}) yield in the extremal states of the spectrum the following uncertainty products:	
	\begin{equation}
	H[\rho_{g. s.}]H[\gamma_{g. s.}]=8\pi e^{2(\gamma-1)}
	\label{eqn:shannongs}
	\end{equation}
	for the ground state, and	
	\begin{equation}
	H[\rho_{Ry}]H[\gamma_{Ry}]\simeq \left(\frac{4\pi}{e^2}\right)^2n
	\label{eqn:shannonry}
	\end{equation}
	for the Rydberg states (i. e. when $n \gg 1$). Remark that these two products fulfil the general entropic uncertainty relation \cite{bbm,bec} which states that
	\begin{equation}
	H[\rho_{n}]H[\gamma_{n}]\ge \pi e
	\label{eqn:shannonineq}
	\end{equation}
	
	Finally, from Eqs. (\ref{eqn:heisenberg}), (\ref{eqn:fisherlengthsineq}) and (\ref{eqn:shannonineq}) we observe that the chain of inequalities
	\begin{equation}
	(\delta x)_n(\delta p)_n<(\Delta x)_n(\Delta p)_n<H[\rho_n]H[\gamma_n]
	\end{equation}
	for both ground and Rydberg states.

\section{Conclusions}

The evaluation of the most relevant spreading and
information-theoretic measures of the half-line Coulomb potential have
been tackled. The position and momentum power and logarithmic moments,
as well as their uncertainty products, have been analytically
computed. Moreover, the logarithmic uncertainty measure and the
information-theoretic lengths of this one-dimensional model have been
analyzed. In particular we have found the exact values of the Fisher
lengths in the two reciprocal spaces as well as the momentum Renyi
lengths. The Shannon length has been also analytically found in
momentum space and for the asymptotic values in position space. The
Fisher length, which quantifies the local disorder of the system,
turns out to be a measure of uncertainty more appropriate than the
quantifiers of global disorder: Heisenberg and Shannon lengths.
Moreover, the position Fisher length of a given physical state is
shown to be directly proportional to the number of nodes of its
corresponding wavefunction and to follow a square-root energy law. 

Contrary to the remaining information measures considered in this
work, the logarithmic uncertainty measure does not depend on the
potential strength $Z$. In addtion, we observe that the uncertainty
product associated to all spreading/information measures here
considered (Heisenberg, logarithmic, Renyi, Onicescu, Shannon and
Fisher) do not depend on $Z$, too. This is in the line shown by
K.D. Sen et. al \cite{pat,sen22,sen23} and other authors \cite{de1,de2} for various simple
quantum-mechanical potentials.

These results allow us to discuss and quantify the internal disorder of the system for both ground and excited states, particularly in the Rydberg region where the classical-quantum transition takes place. Finally, let us point out that we are now in the best position to compute the complexity measures of this physical model, so much used nowadays in a wide range of scientific and technological areas (see  e. g., \cite{c02,a08a,s08}). To do that in an analytical way, however, it is required the explicit evaluation of the functionals of the Laguerre polynomials involved in the expressions (\ref{eqn:nupos}), (\ref{eqn:nuq2pos}) and (\ref{eqn:shannonrhox})-(\ref{eqn:eshannonrhox}) of the entropic moments and the Shannon entropy of the system in position space, what is not a trivial task. This work has been recently accomplished \cite{psmo}.

\section*{Acknowledgements}
This work has been partially supported by the Spanish MICINN grant FIS2008-02380, and the grants FQM-1735 and FQM-2445 of Junta de Andalucía. The authors belong to the research group FQM-207. J.J. Omiste acknowledges the scholarship ``Iniciaci\'on a la investigaci\'on'' in the framework of the Plan Propio of the University of Granada.


\begin{thebibliography}{0}

\bibitem{cas}
G. Casati, B. V. Chirikov, D. L. Shepelyansky \& I. Guarneri, Relevance of classical chaos in quantum mechanics: The hydrogen atom in a monochromatic field, Phys. Rep. 154 (1987) 77-123; IEEE J. Quantum Electron. QE-24 (1988) 1420-1444

\bibitem{may}
M. Mayle et al, One-dimensional Rydberg gas, Phys. Rev. Lett. 99 (2007) 113004

\bibitem{leo}
J. G. Leopold \& I. C. Percival, Microwave Ionization and Excitation of Rydberg Atoms, Phys. Rev. Lett. 41 (1978) 944; D. Richards, Ionisation of excited one-dimensional hydrogen atoms by low-frequency fields, J. Phys. B 20 (1987) 2171; J. G. Leopold \& D. Richards, A study of quantum dynamics in the classically chaotic regime, ibid 21 (1988) 2179

\bibitem{sto}
C. L. Stokey et al, Production of quasi-one-dimensional very-high-n Rydberg atoms, Phys. Rev. A 67 (2003) 013403

\bibitem{pen}
V. L. Pen \& T. F. Jiang, Strong-field effects of the one-dimensional hydrogen atom in momentum space, Phys. Rev. A 46 (1992) 4297-4305

\bibitem{nie}
M. M. Nieto, Electrons above a helium surface and the one-dimensional Rydberg atom, Phys. Rev. A 61 (2000) 034901

\bibitem{jen}
R. V. Jensen, Stochastic Ionization of Surface-State Electrons Phys. Rev. Lett. 49 (1982) 1365; Stochastic ionization of surface-state electrons: Classical theory, Phys. Rev. A 30 (1984) 386-397

\bibitem{dyk}
M. I. Dykman, P. M. Playzman \& P. Seddigrad, Qubits with electrons on liquid helium, Phys. Rev. B 67 (2003) 155402

\bibitem{pla}
P. M. Platzman \& M. I. Dykman, Quantum computing with electrons on liquid helium, Science 284 (1999) 1967

\bibitem{jak}
D. Jaksch et al, Fast quantum gates for neutral atoms, Phys. Rev. Lett. 85 (2000) 2208

\bibitem{vei}
R. Veilande \& I. Bersons, Wave packet fractional revivals in a one-dimensional Rydberg atom, J. Phys. B40 (2007) 2111-2119

\bibitem{fis}
W. Fischer, H. Leschke \& P. Müller, The functional-analytic versus the functional-integral approach to quantum Hamiltonians: The one-dimensional hydrogen atom. J. Math. Phys. 36 (1995) 2313

\bibitem{gui}
T. Guillot, A comparison of the interior of Jupiter and Saturn, Planet. Space Sci. 47 (1999) 1183

\bibitem{ave}
J. Avery, Hyperspherical Harmonics: Applications in Quantum Theory, Kluwer Academic, Dodrecht, (1988)

\bibitem{si1}
H. S. Sichel, Fitting growth and frequency curves by the method of frequency moments, J. Roy. Statist. Soc. 110 (1947) 337-347.

\bibitem{yul}
G. U. Yule, On some properties of normal distributions, univariate and bivariate, based on sums of squares of frequencies, Biometrika 30 (1938) 1-10

\bibitem{ken}
M. G. Kendall and A. Stuart, The Advanced Theory of Statistics, vol. 1 (Charles Griffin Co., London, 1969)

\bibitem{si2}
H. S. Sichel, The method of frequency-moments and its applications to type VII Populations, Biometrika 36 (1949) 404

\bibitem{she}
L. R. Shenton, Efficiency of the method of moments and the Gram-Charlier type A distribution, Biometrika 38 (1951) 58-73

\bibitem{rom}
E. Romera, J. C. Angulo and J. S. Dehesa, Reconstruction of a density from its entropic moments, in R. L. Fry (ed.), The 21st. International Workshop on Bayesian Inference and Maximum Entropy Methods in Science and Engineering (A. I. P., New York, 2002), pp. 449-457

\bibitem{uff}
J. B. M. Uffink, Measures of uncertainty and the uncertainty principle, Ph. D. Thesis, University of Utrech, 1990

\bibitem{ha1}
M. J. W. Hall, Universal geometric approach to uncertainty entropy and information, Phys. Rev. A59 (1999) 2602-2615

\bibitem{ha2}
M. J. W. Hall, Exact uncertainty measures, Phys. Rev. A 64 (2001) 052103

\bibitem{hel}
O. Onicescu, Energie informationalle, C.R. Acad. Sci. Paris. A263 (1966) 841; E. Heller, Quantum localization and the rate of exploration of phase space, Phys. Rev. A 35 (1987) 1360

\bibitem{de1}
P. Sánchez-Moreno, R. González-Férez and J. S. Dehesa, Improvement of the Heisenberg and Fisher-information-based uncertainty relations for D-dimensional central potentials, New J. Phys. 8 (2006) 330

\bibitem{de2}
J. S. Dehesa, R. González-Férez and P. Sánchez-Moreno, The Fisher-information based uncertainty relation, Cramer-Rao inequality and kinetic energy for the D-dimensional central problem, J. Phys. A 40 (2007) 1845-1856

\bibitem{pat}
S. H. Patil, K. D. Sen, N. A. Watson \& H. E. Montogomery Jr., Characteristic measures of net information measures for constrained Coulomb potentials, J. Phys. B 40 (2007) 2147-2162

\bibitem{san}
J. Sánchez-Ruiz, J. S. Dehesa, Entropic integrals of orthogonal hypergeometric polynomials with general supports, J. Comp. Appl. Math. 118 (2000) 311-322

\bibitem{dya}
J. S. Dehesa, R. J. Yáñez, A. I. Aptekarev and V. Buyarov, Strong asymptotics of Laguerre polynomials and information entropies of two-dimensional harmonic oscillator and one-dimensional Coulomb potentials, J. Math. Phys. 39 (1998) 3050

\bibitem{rai}
U. Raitzsch et al, An echo experiment in a strongly interacting Rydberg gas, Phys. Rev. Lett. 100, p. 013002 (2008)

\bibitem{kau}
B. Kaulakis, Quasiclassical dipole matrix elements for high atomic states and stochastic dynamics of hydrogen atoms in microwave fields, J. Phys. B 24 (1991) 571-585 

\bibitem{bbm}
I. Bialynicki-Birula and J. Mycielski, Uncertainty relations for information entropy, Commun. Math. Phys. 44 (1975) 129

\bibitem{bec}
W. Beckner, Pitt's inequality and the uncertainty principle, Proc. Amer. Math. Soc. 123 (1995) 1897

\bibitem{c02}
R. G. Catalán, J. Garay \& R. López-Ruiz, Features of the extension of a statistical measure of complexity to continuous systems, Phys. Rev. E66 (2002) 011102

\bibitem{a08a}
J. C. Angulo \& J. Antolín, Atomic complexity measures in position and momentum spaces, J. Chem. Phys. 128 (2008) 164109

\bibitem{s08}
K. D. Sen, J. Antolín \& J. C. Angulo, Fisher-Shannon analysis of ionizaton processes and isoelectronic series, Phys. Rev A76 (2007) 032502

\bibitem{sen22} S.H. Patil \& K.D. Sen, Scaling properties of net information measures for superpositions of power potentials: Free and spherically confined cases, Phys. Lett. A 370 (2007) 354

\bibitem{sen23} K.D. Sen \& J. Katriel, Information entropies for eigendensities of homogeneous potentials, J. Chem. Phys. 125 (2006) 074117

\bibitem{psmo} P. S\'anchez-Moreno, J.J. Omiste and J.S. Dehesa, Entropic functionals of Laguerre polynomials and complexity properties of the half-line Coulomb potential, Preprint (2009)

\end{thebibliography}
\end{document}